\documentclass{ws-ijmpb}
\begin{document}

\markboth{A.P.Alodjants, S.M.Arakelian} {Quantum storage and
cloning of light states in EIT-like medium}

%
\catchline{}{}{}{}{}
%

\title{QUANTUM STORAGE AND CLONING OF LIGHT\\
STATES IN EIT-LIKE MEDIUM}

\author{A.P. ALODJANTS}

\address{Department of Physics and Applied Mathematics,\\ Vladimir State University, Gorkogo str., 87\\
Vladimir, 600000, Russia\\ alodjants@vpti.vladimir.ru }

\author{S.M. ARAKELIAN}

\address{Department of Physics and Applied Mathematics,\\ Vladimir State University, Gorkogo str., 87\\
Vladimir, 600000, Russia\\ arak@vpti.vladimir.ru }

\maketitle

\begin{history}
\received{Day Month Year} \revised{Day Month Year}
\end{history}

\begin{abstract}
In the paper we consider a new approach for storage and cloning of
quantum information by three level atomic (molecular) systems in
the presence of the electromagnetically induced transparency (EIT)
effect. For that, the various schemes of transformation into the
bright and dark polaritons for quantum states of optical field in
the medium are proposed. Physical conditions of realization of
quantum nondemolition (QND) storage of quantum optical state are
formulated for the first time. We have shown that the best storage
and cloning of  can be achieved with the atomic ensemble in the
Bose-Einstein condensation state. We discuss stimulated Raman
two-color photoassociation for experimental realization of the
schemes under consideration.
\end{abstract}


\section{Introduction}

Modern development of quantum information, communication and
especially quantum cryptography requires the new methods and
approaches in quantum information processing. A significant
progress on this way can be achieved with the help of novel
devices for quantum storage, memory and transmission of
information embodied in optical continuous variables, i.e. in
Hermitian quadratures\cite{julsgaard}. Typically, the memory
devices proposed now for those purposes explore various methods of
entanglement and mapping of quantum state of light onto the atomic
ensembles\cite{kozhekin}$^{-}$\cite{liu}.

One practically important and interesting possibility to store
optical information is connected with the usage of effect of
electromagnetically induced transparency (EIT) taking place in
three level atomic systems with $\Lambda $-configuration of the
energy levels - see Fig.1. In this case strong (classical)
coupling field (i.e. control field) creates transparency window in
the medium, and second weak (quantum) probe field (we call it as a
signal) propagates through the resonant atomic system with a very
small absorption\cite{fleischauer}$^-$\cite{prokhorov1}. The EIT
effect is accompanied with significant reduction of observable
group velocity of signal pulse, as well. In Ref.~\refcite{liu} the
authors propose to use such a ``stopped'' light for coherent
(classical) storage of information in ultracold ($T \simeq 450$~
nK) sodium atoms near the transition to the Bose-Einstein
condensation (BEC) state. In fact, the phenomenon is displayed by
switching of signal pulse with delay time $\tau \simeq 45\; \mu
{\rm sec}$ and more. The Doppler effect, being limitation for the
process, can be suppressed for the case. Physically, quantum
properties of the coupling system, i.e. signal optical pulse and
atomic system under the EIT condition, can be described in terms
of ``bright'' and ``dark'' polaritons for atom-field excitations -
see Refs.~\refcite{fleischauer},~\refcite{prokhorov1} for more
details. The experiment has been carried out in
Ref.~\refcite{phillips} for hot ($T \simeq 360$~K) atomic Rb vapor
cell, and quantum state transfers from the light to atomic
excitations.

Another approach to store and transmit quantum information by
memory devices has been established in Ref.~\refcite{julsgaard}.
In particular, the fidelity up to $70\% $ and memory lifetime of
up to $4\;{\rm msec}$ has been achieved experimentally for
Gaussian light pulses using complex (three step passing) scheme
for light-atom interaction and subsequent measurements by feedback
introduced onto the atoms.

In present paper we consider also the problem of storage of
quantum state of optical field by using the quantum cloning
procedure for continuous variables for the light-atom
interaction\cite{fiurasek}. It is
shown\cite{braunstein}$^-$\cite{grosshans} that although perfect
cloning is forbidden in quantum theory the fidelity of such a
process can be high enough and equal $2/3$.

In Section 2 we propose quantum non-demolition (QND) method to
store the quantum state of light what is very closed to QND
measurement procedure in quantum and atomic
optics\cite{braginsky}$^,$~\cite{alodjants}. In this case the
specific properties of quantum state of storage device become
important. Necessary criteria are formulated for estimation of
efficiency of optical field storage and transmission. In Section 3
we consider a new scheme for optimal quantum cloning of optical
field in the EIT-like system. In the case a preliminary linear
amplification of the signal optical pulse is absolutely necessary.
In conclusion we briefly discuss appropriate application of the
schemes proposed in the paper for the problem of quantum
cryptography with continuous variables. In Appendix we also
discuss an implementation of the stimulated two-photon Raman
photoassociation\cite{drummond}$^,$~\cite{winkler} effect under
the BEC condition as a tool for experimental realization of the
problem under consideration.

\section{Quantum non-demolition storage of the state of optical field}

Let us consider interaction of three level atomic system with two
optical fields under the three effectively interacting quantum
modes approximation. In this regime all atoms macroscopically
occupy three internal levels depicted in Fig.1. The three-mode
approach is valid when the spatial state (motional degree of
freedom) of atomic cloud still unchanged on the time of atom-field
interaction and also still essentially independent on internal
atomic states. For ultracold atoms near the condensation
transition such a conditions are fulfilled for small number of
particles ($N \lesssim 10^4$) or relevant size of medium and trap
size (spacing between the trap modes $\hbar \omega_{trap}$ that is
numerically corresponds to frequency $\omega_{trap}\le 100 {\rm
Hz}$)~\cite{anglin}. The macroscopic atomic coherence can be also
achieved for hot atomic clouds. There exist some theoretical and
experimental proposals for that -- see e.g. Ref.~\refcite{adams}.
In particular, the atomic system that is described by collective
magnetic momentum $J$ of ground state is especially have been
prepared in coherent spin state for which only one projection, say
$J_x$, have non-zero average value for current
experiments\cite{julsgaard}$^,$\cite{kozhekin}. Another
possibility, what is important for us in the paper, is based on
achieving of EIT regime for atom-field interaction\cite{phillips}.

Thus, we will describe the bright (light) and dark (matter)
polaritons in EIT-like medium (being the result of excitation of
quantum coupled state for optical field and matter) by unitary
linear transformation for two annihilation operators $\Phi _{f} $
and $\Phi _{\xi}  $, respectively, in the form:

\begin{equation}
\label{eq1}
\Phi _{f} = \mu f_{in} - \nu \xi _{in} ,
\quad
\Phi _{\xi}  = \mu \xi _{in} + \nu f_{in} ,
\end{equation}

\noindent where $f_{in} $ is the annihilation operator for signal
(probe) field at the input of atomic medium. In the Schwinger
bosons representation\cite{levy} spin operator $\xi = {{a_{2}^{ +}
a_{1}} \mathord{\left/ {\vphantom {{a_{2}^{ +} a_{1}}  {\sqrt {N}}
}} \right. \kern-\nulldelimiterspace} {\sqrt {N}} }$ characterizes
macroscopic excitations in atomic system, $a_{j} $ ($a_{j}^{ +} $)
is the atom annihilation (creation) operator for $j$-th internal
level ($j = 1,2$) respectively, $N$ is the total number of
particles. The parameters $\mu = \cos\left( {\theta} \right)$ and
$\nu = \sin\left( {\theta}  \right)$ in Eq.~(\ref{eq1}) depend on
oblique angle $\theta $ between signal and control field. The
value $\theta $ defines the efficiency of linear transformation
(\ref{eq1}) and energy exchange for polaritons. In the presence of
the EIT effect it can be represented as (see
Ref.~\refcite{fleischauer}):

\begin{equation}
\label{eq2} \frac{{\nu ^{2}}}{{\mu ^{2}}} = \tan^{2}\left(
{\theta}  \right) = \frac{{\Omega _{2}^{2} N}}{{\Omega _{1}^{2}} }
= n_{g} ,
\end{equation}

\noindent where $\Omega _{j} $ ($j = 1,2$) characterizes the Rabi
frequency for control and signal fields respectively, $n_{g} $ is
the refractive index.

\begin{figure}[bt]
\centerline{\psfig{file=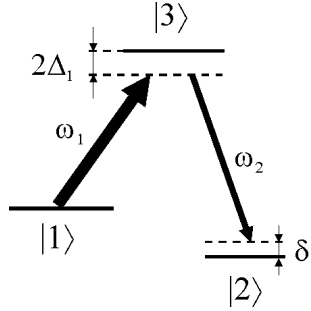,width=4cm}} \vspace*{8pt}
\caption{Energy levels for $\Lambda $-scheme under the EIT effect;
$\omega _{1} $ ($\omega _{2} $) is the frequency of contol
(signal) field, $2\Delta _{1} $ and $\delta $ are the detunings
from the exact resonance.}
\end{figure}

In adiabatic approximation we assume that the atoms are populated
mostly at the lowest (ground) level $\left| {2} \right\rangle $,
and following inequalities are valid for average number of atoms
$n_{j} = \left\langle {a_{j}^{ +}  a_{j}}  \right\rangle $ ($j =
1,2,3$) - see also Fig.1:

\begin{equation}
\label{eq3}
n_{3} \ll n_{1} < n_{2} .
\end{equation}

With the help of expressions (\ref{eq1}) for quadrature components
of polaritons $Q_{j}^{out} = \Phi _{j} + \Phi _{j}^{ +}  $,
$P_{j}^{out} = i\left( {\Phi _{j}^{ +}  - \Phi _{j}}  \right)$, $j
= f,\xi $ one can obtain for output of the system:

\begin{eqnarray}
\nonumber \label{eq4} Q_{f}^{out} = \mu Q_{f}^{in} - \nu Q_{\xi}
^{in} , \quad P_{f}^{out} = \mu P_{f}^{in} - \nu P_{\xi} ^{in} ,\\
Q_{\xi} ^{out} = \mu Q_{\xi} ^{in} + \nu Q_{f}^{in} ,
\quad
P_{\xi} ^{out} = \mu P_{\xi} ^{in} + \nu P_{f}^{in} ,
\end{eqnarray}

\noindent
where quadratures $Q_{f}^{in} = f_{in} + f_{in}^{ +}  $, $P_{f}^{in} =
i\left( {f_{in}^{ +}  - f_{in}}  \right)$ relate to signal field at the
input of the medium. The values $Q_{\xi} ^{in} = \xi _{in} + \xi _{in}^{ +}
\equiv {{\left( {a_{1}^{ +}  a_{2} + a_{2}^{ +}  a_{1}}  \right)}
\mathord{\left/ {\vphantom {{\left( {a_{1}^{ +}  a_{2} + a_{2}^{ +}  a_{1}}
\right)} {\sqrt {N}} }} \right. \kern-\nulldelimiterspace} {\sqrt {N}} }$
and $P_{\xi} ^{in} = i\left( {\xi _{in}^{ +}  - \xi _{in}}  \right) \equiv
{{i\left( {a_{1}^{ +}  a_{2} - a_{2}^{ +}  a_{1}}  \right)} \mathord{\left/
{\vphantom {{i\left( {a_{1}^{ +}  a_{2} - a_{2}^{ +}  a_{1}}  \right)}
{\sqrt {N}} }} \right. \kern-\nulldelimiterspace} {\sqrt {N}} }$ describe
the elementary excitations (i.e. the spin wave components) and satisfy the
standard SU(\ref{eq2})-algebra commutation relations.$^{} $

We consider, further, the quantum non-demolition (QND) storage of
information by signal field quadratures $Q_{f}^{in} $ and
$P_{f}^{in} $ measurement. The procedure is shown in Fig.2: the
measurement of polariton quadratures $Q_{\xi} ^{out} $ and
$P_{\xi} ^{out} $ at the output of the device gives information on
the values of signal quadratures $Q_{f}^{in} $ and $P_{f}^{in} $
at the input.

Let us introduce following correlation coefficients to describe
the process under consideration in Heisenberg representation (cf.
Ref.~\refcite{alodjants}):

\begin{equation}
\label{eq6} C_{1}^{\left( {X} \right)} = \frac{{\left|
{\left\langle {X_{f}^{in} X_{f}^{out}}  \right\rangle -
\left\langle {X_{f}^{in}}  \right\rangle \left\langle
{X_{f}^{out}}  \right\rangle}  \right|}}{{V_{X,f}^{in}
V_{X,f}^{out}} }^{2},
\end{equation}

\begin{equation}
\label{eq7} C_{2}^{\left( {X} \right)} = \frac{{\left|
{\left\langle {X_{f}^{in} X_{\xi }^{out}}  \right\rangle -
\left\langle {X_{f}^{in}}  \right\rangle \left\langle {X_{\xi}
^{out}}  \right\rangle}  \right|}}{{V_{X,f}^{in} V_{X,\xi} ^{out}}
}^{2},
\end{equation}

\begin{equation}
\label{eq8}
C_{3}^{\left( {X} \right)} = \frac{{\left| {\left\langle {X_{f}^{out} X_{\xi
}^{out}}  \right\rangle - \left\langle {X_{f}^{out}}  \right\rangle
\left\langle {X_{\xi} ^{out}}  \right\rangle}  \right|}}{{V_{X,f}^{out}
V_{X,\xi} ^{out}} }^{2},
\end{equation}

\noindent where continuous variable $X = \left\{ {Q,\left. {P}
\right\}} \right.$; $V_{X,f}^{in\left( {out} \right)} =
\left\langle {\left( {\Delta X_{f}^{in\left( {out} \right)}}
\right)^{2}} \right\rangle $, $V_{X,\xi }^{in\left( {out} \right)}
= \left\langle {\left( {\Delta X_{\xi} ^{in\left( {out} \right)}}
\right)^{2}} \right\rangle $ are the variances of quadrature
components for signal pulse and polariton, respectively, before
(after) storage procedure.

\begin{figure}[bt]
\centerline{\psfig{file=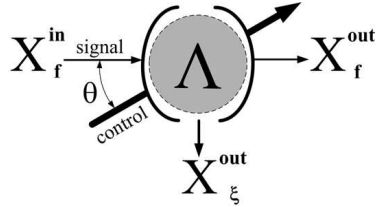,width=5cm}} \vspace*{8pt}
\caption{Scheme of the QND storage of the $X_{f}^{in} $
-quadratures ($X = \left\{ {Q,\left. {P} \right\}} \right.$) for
signal field in the EIT-like medium ($\Lambda $); $\theta $ is the
oblique angle between control and signal fields. The $X_{f,\xi}
^{out} $ are the polariton quadratures at the output of the
medium.}
\end{figure}

The conditions (\ref{eq6})--(\ref{eq8}) take into account the
correlations for two quadratures $Q_{f}^{in} $ and $P_{f}^{in} $
simultaneously (in contrast with the quantum non-demolition
measurement procedure for a single quadrature component
$Q_{f}^{in} $ or $P_{f}^{in} $ (see e.g.
Ref.~\refcite{braginsky}), and have a simple physical meaning. The
correlation coefficient $C_{1}^{\left( {Q,P} \right)} $ takes into
account the degree of distortion for quadratures $Q_{f}^{in} $ and
$P_{f}^{in} $ of input signal at the output of the read-out
device. The ``quality'' of non-demolition recording of optical
information for that can be established by using the
$C_{2}^{\left( {Q,P} \right)} $ coefficients. The ability of the
storage device to prepare the quantum state for signal field in
respect of further transmission is characterized by the
$C_{3}^{\left( {Q,P} \right)} $ coefficients. In other words, the
coefficients $C_{3}^{\left( {Q,P} \right)} $ can be associated
with the cloning procedure for signal pulse into the considered
atomic ensemble. In fact, the coefficients in Eq.~(\ref{eq8})
represent a necessary condition for such a procedure - see also
Eq.~(\ref{eq21a}) below.

In the case of the ideal QND-storage the all correlation
coefficients are equal to one, i.e. $C_{j}^{\left( {X} \right)} =
1$, $X = Q,P$, $j = 1,2,3$. The fact occurs when

\begin{eqnarray}
\label{eq9} \nonumber Q_{f}^{out} = Q_{f}^{in} , \quad Q_{\xi}
^{out} = GQ_{f}^{in} \\
P_{f}^{out} = P_{f}^{in} ,
\quad
P_{\xi} ^{out} = GP_{f}^{in} ,
\end{eqnarray}

\noindent where $G$ is QND gain. In the case of ideal cloning
procedure the value $G = 1$ - cf.~(\ref{eq21a}).

From expressions (\ref{eq9}) and Eqs.~(\ref{eq4}) it is easy to
see that the EIT-like medium is not ideal for quantum information
storage but in some cases can be closed to that. Actually, using
the Eqs.~(\ref{eq4}), we can rewrite the expressions
(\ref{eq6})--(\ref{eq8}) in the form:

\begin{eqnarray}
\label{eq11} \nonumber C_{1}^{\left( {X} \right)} = \frac{{\mu
^{2}V_{X,f}^{in}} }{{\mu ^{2}V_{X,f}^{in} + \nu ^{2}V_{X,\xi}
^{in}} }, \quad C_{2}^{\left( {X} \right)} = \frac{{\nu
^{2}V_{X,f}^{in}} }{{\mu ^{2}V_{X,\xi} ^{in} + \nu
^{2}V_{X,f}^{in}} }, \\ C_{3}^{\left( {X} \right)} = \frac{{\mu
^{2}\nu ^{2}\left( {V_{X,f}^{in} - V_{X,\xi} ^{in}}
\right)^{2}}}{{\left( {\mu ^{2}V_{X,f}^{in} + \nu ^{2}V_{X,\xi}
^{in}}  \right)\left( {\mu ^{2}V_{X,\xi} ^{in} + \nu
^{2}V_{X,f}^{in}}  \right)}},
\end{eqnarray}

\noindent
where $X = Q,P$.

From expressions (\ref{eq11}) follow that correlation coefficients
$C_{j}^{\left( {X} \right)}$ can satisfy simultaneously to
condition $C_{j}^{\left( {X} \right)} = 1$ ($j = 1,2,3$) only for
the case when all variances $V_{Q,\xi }^{in} = V_{P,\xi} ^{in} =
0$. The condition corresponds to initially prepared perfect
spin-sqeezed state for atomic
medium\cite{kozhekin}$^,$\cite{cirac2}. If it is not the case, the
quality of the storage and transmission of information depends on
quantum properties of atomic system, and also is determined by
oblique angle $\theta $.

In particular, the coefficient $C_{1}^{\left( {X} \right)} \simeq
1$ for $\left| {\nu}  \right| \ll 1$ ($\theta \ll 1$) in the limit
of the non-demolition storage of signal field quadratures. When
$C_{2,3}^{\left( {X} \right)} \to 1$, as well, we must require the
following inequalities to be fulfilled for the variances:

\begin{equation}
\label{eq13a} V_{Q,\xi} ^{in} \ll \frac{{\nu ^{2}}}{{\mu ^{2}}},
\end{equation}
\begin{equation}
\label{eq13b} V_{P,\xi} ^{in} \ll \frac{{\nu ^{2}}}{{\mu ^{2}}}.
\end{equation}

Note, that the conditions (\ref{eq13a}), (\ref{eq13b}) are not
require in essential suppression of group velocity for signal
pulse passing the atomic medium - see also (\ref{eq2}).

Let us examine the criteria (\ref{eq13a}), (\ref{eq13b}) under the
conditions of the experiments related to the EIT-effect
observation and carried out recently\cite{liu}$^,$\cite{phillips}.
We assume the probe field is in arbitrary coherent state with the
variances $V_{X,f}^{in} = 1$, for simplicity.

For two-mode Fock state $\left| {\psi}  \right\rangle _{at} =
\left| {n_{1} } \right\rangle \left| {n_{2}}  \right\rangle $ of
atomic system under condition (\ref{eq3}) we readily find from
expressions (\ref{eq11}):

\begin{equation}
\label{eq14} C_{1}^{\left( {X} \right)} \simeq \frac{{\mu
^{2}}}{{1 + 2\nu ^{2}n_{1}} }, \quad C_{2}^{\left( {X} \right)}
\simeq \frac{{\nu ^{2}}}{{1 + 2\mu ^{2}n_{1}} }, \quad
C_{3}^{\left( {X} \right)} \simeq \frac{{4\mu ^{2}\nu
^{2}n_{1}^{2}} }{{1 + 4\mu ^{2}\nu ^{2}n_{1}^{2} + 2n_{1}} }.
\end{equation}

The maximal values for coefficients $C_{j}^{\left( {X} \right)} $
can be achieved in the limit of $n_{1} \ll 1$. The case
corresponds to coherent atomic medium with the variance of
polaritons $V_{X,\xi} ^{in} = 1$. Then, one can obtain from
Eqs.~(\ref{eq11}), (\ref{eq14}):

\begin{equation}
\label{eq15} C_{1,class}^{\left( {X} \right)} \simeq \mu ^{2},
\quad C_{2,class}^{\left( {X} \right)} \simeq \nu ^{2}, \quad
C_{3,class}^{\left( {X} \right)} = 0.
\end{equation}

The relations (\ref{eq15}) for correlation coefficients
(\ref{eq6})--(\ref{eq8}) determine an ultimate case of classical
recording of information by coherent atomic device when the
conditions (\ref{eq13a}), (\ref{eq13b}) are violated.

For the Bose-Einstein condensate medium the initially prepared state

\begin{equation}
\label{eq16}
\left| {\psi}  \right\rangle _{BEC} = \frac{{1}}{{\sqrt {N!}} }\left(
{\alpha _{1} a_{1}^{ +}  + \alpha _{2} a_{2}^{ +} }  \right)^{N}\left| {0}
\right\rangle
\end{equation}

\noindent is entangled\cite{prokhorov2}$^,$\cite{cirac2}. The
parameters $|\alpha_j|^2=\left. n_j \right/ N$ ($j = 1,2$)
determine a relative atomic population at the levels $\left| {1}
\right\rangle $ and $\left| {2} \right\rangle $, respectively, and
obey normalization condition $\left| {\alpha _{1}} \right|^{2} +
\left| {\alpha _{2}} \right|^{2} = 1$. In this case we have for
the variances $V_{X,\xi} ^{in} $:

\begin{equation}
\label{eq17} V_{Q,\xi} ^{in} = 1 - 4\left| {\alpha _{1}}
\right|^{2}\left| {\alpha _{2} } \right|^{2}\cos^{2}\left(
{\varphi}  \right), \quad V_{P,\xi} ^{in} = 1 - 4\left| {\alpha
_{1}}  \right|^{2}\left| {\alpha _{2} } \right|^{2}\sin^{2}\left(
{\varphi}  \right),
\end{equation}

\noindent
where $\varphi = \varphi _{2} - \varphi _{1} $ is relative atomic phase
($\alpha _{1,2} = \left| {\alpha _{1,2}}  \right|e^{i\varphi _{1,2}} $).

As follows from relations (\ref{eq17}), the variances $V_{Q,\xi}
^{in} $ and $V_{P,\xi} ^{in} $ are in spin-squeezed states
simultaneously, and the values $V_{Q,\xi} ^{in} = V_{P,\xi} ^{in}
= {{1} \mathord{\left/ {\vphantom {{1} {2}}} \right.
\kern-\nulldelimiterspace} {2}}$ are for $\left| {\alpha _{1} }
\right|^{2} = \left| {\alpha _{2}}  \right|^{2} = {{1}
\mathord{\left/ {\vphantom {{1} {2}}} \right.
\kern-\nulldelimiterspace} {2}}$, $\varphi = {{\pi}
\mathord{\left/ {\vphantom {{\pi}  {4}}} \right.
\kern-\nulldelimiterspace} {4}}$. In this case the relevant
coefficients (see Eqs.~(\ref{eq11})) are maximal $C_{1,2}^{\left(
{X} \right)} = {{2} \mathord{\left/ {\vphantom {{2} {3}}} \right.
\kern-\nulldelimiterspace} {3}}$, $C_{3}^{\left( {X} \right)} =
{{1} \mathord{\left/ {\vphantom {{1} {9}}} \right.
\kern-\nulldelimiterspace} {9}}$, and we can achieve essentially
quantum storage of information. The conditions (\ref{eq13a}) and
(\ref{eq13b}) are fulfilled simultaneously for the value of angle
$\theta \simeq {{\pi} \mathord{\left/ {\vphantom {{\pi} {4}}}
\right. \kern-\nulldelimiterspace} {4}}$.

The obtained dependences for correlation coefficients $C_{j}^{} \equiv
C_{j}^{\left( {Q} \right)} $ ($j = 1,2,3$) vs oblique angle $\theta $ are
plotted in Fig.3.

\begin{figure}[bt]
\centerline{\psfig{file=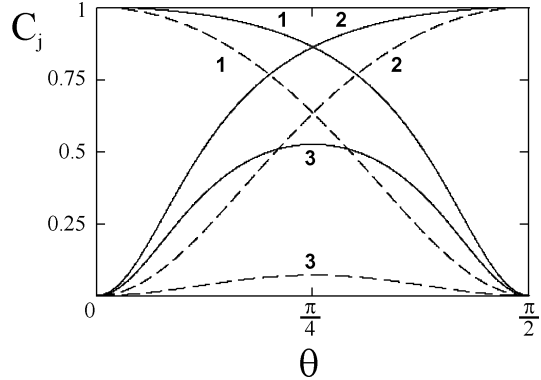,width=7cm}} \vspace*{8pt}
\caption{Dependences of correlation coefficients $C_{j}^{} $ on
the $\theta $ angle; $j=1$- curves 1, $j=2$- curves 2, and $j=3$ -
curves 3. The magnitudes of parameters used for calculations are:
$\left| {\alpha _{1}}  \right|^{2} = 0.3$, $\varphi = \pi/4$
(dashed lines) and $\varphi = 0$ (solid lines) respectively.}
\end{figure}

The variances $V_{Q,\xi} ^{in} $ correspond to squeezed state for
the relative phase $\varphi = 0$, and the condition (\ref{eq13a})
is fulfilled only. In this regime EIT-medium performs quantum
non-demolition measurement of the $Q_{f}^{in} $ quadrature only.
At the same time, for another quadrature component $P_{f}^{in} $
the variance corresponds to coherent level of fluctuations when
$V_{P,\xi} ^{in} = 1$ and $C_{j}^{\left( {P} \right)} =
C_{j,class} $ respectively - see Eq.(\ref{eq15}). Practically,
such a method of the quantum information storage has a narrower
area of applications due to dependence on signal field quadratures
$Q_{f}^{in} $ and $P_{f}^{in} $.

\section{Quantum cloning with polaritons}

Now we switch our attention to the problem of cloning of optical
field $f_{in} $ onto the ultracold atomic (molecular) ensemble
with the help of the EIT-like medium. The principal set-up for
optimal cloning procedure under consideration is shown in Fig.4.

The initial radiation propagates through the phase-insensitive
linear amplifier (LA) with the gain that is equal two. The linear
transformation for photon annihilation operators of signal ($f_{}
$) and ancilla ($c$) modes for symmetric cloning is represented in
the form:

\begin{equation}
\label{eq18}
f = \sqrt {2} f_{in} + c_{in}^{ +}  ,
\quad
c_{out} = \sqrt {2} c_{in}^{} + f_{in}^{ +}  ,
\end{equation}

\noindent where $c_{in} $ ($c_{out} $) is the value of operator at
the input (output) of the amplifier (we assume that $c_{in} $ is
in the vacuum state).

\begin{figure}[bt]
\centerline{\psfig{file=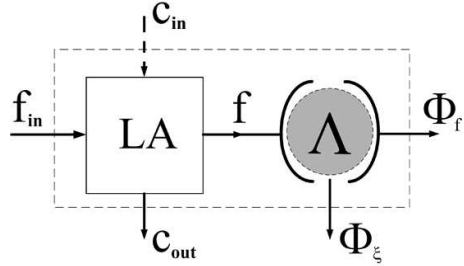,width=6cm}} \vspace*{8pt}
\caption{Scheme of quantum cloning for light onto the polaritons.
Here $f_{in} $ ($c_{in}^{} $) is annihilation operator for signal
(ancilla) optical mode, respectively at the input of cloning
device; LA is the linear amplifier, $\Lambda $ is the EIT-like
medium, $\Phi _{f} $ and $\Phi _{\xi}  $ are the polariton
operators at the output of apparatus.}
\end{figure}

Then, according to the scheme in Fig.4, the signal optical field
is stored in the EIT medium. The bright and dark polaritons at the
output of medium represent two symmetric clones of initial field
$f_{in} $ as a result.

Taking into account the Eqs.~(\ref{eq1}) and Eqs.~(\ref{eq18})
with $\theta \simeq {{\pi} \mathord{\left/ {\vphantom {{\pi}
{4}}} \right. \kern-\nulldelimiterspace} {4}}$ we represent a
total unitary transformation for annihilation operators of bright
(optical) $\Phi _{f} $ and dark (excitation in matter) $\Phi _{\xi
} $ clones as:

\begin{equation}
\label{eq19}
\Phi _{f} = f_{in} + \frac{{1}}{{\sqrt {2}} }\left( {c_{in}^{ +}  - \xi
_{in}}  \right),
\quad
\Phi _{\xi}  = f_{in} + \frac{{1}}{{\sqrt {2}} }\left( {c_{in}^{ +}  + \xi
_{in}}  \right).
\end{equation}

Corresponding to Eqs.~(\ref{eq19}) the quadrature component
transformations have the form:

\begin{equation}
\label{eq20}
Q_{f,\xi} ^{out} = Q_{f}^{in} + \frac{{1}}{{\sqrt {2}} }\left( {Q_{c}^{in}
\mp Q_{\xi} ^{in}}  \right),
\quad
P_{f,\xi} ^{out} = P_{f}^{in} - \frac{{1}}{{\sqrt {2}} }\left( {P_{c}^{in}
\pm P_{\xi} ^{in}}  \right),
\end{equation}

\noindent
where $Q_{c}^{in} = c_{in} + c_{in}^{ +}  $, $P_{c}^{in} = i\left( {c_{in}^{
+}  - c_{in}}  \right)$ are the Hermirtian quadratures for ancilla vacuum
mode at the input of amplifier (see Fig.4).

The expressions (\ref{eq19}), (\ref{eq20}) represent the desired
linear transformations for optimal cloning procedure of continuous
variables in the Heisenberg picture\cite{braunstein}. In the case,
corresponding mean values of quadratures (\ref{eq20}) and their
variances fulfill the expressions (cf.~(\ref{eq9})):

\begin{equation}
\label{eq21a} \left\langle {X_{f,\xi} ^{out}}  \right\rangle =
\left\langle {X_{f}^{in}} \right\rangle ,
\end{equation}
\begin{equation}
\label{eq21b} V_{X}^{out}\equiv V_{X,f}^{out} = V_{X,\xi}^{out} =
V_{X,f}^{in} + 1,
\end{equation}

\noindent where variable $X = \left\{ {Q,P} \right\}$.

Last term in Eq.(16b) characterizes the impossibility to perfect
cloning of quantum state. For the relevant correlation
coefficients (\ref{eq6})--(\ref{eq8}) we have in this case:

\begin{equation}
\label{eq22}
C_{1,clone}^{\left( {X} \right)} = C_{2,clone}^{\left( {X} \right)} =
\frac{{1}}{{2}},
\quad
C_{3,clone}^{\left( {X} \right)} = \frac{{1}}{{4}}.
\end{equation}

Two coefficients $C_{1,2}^{\left( {X} \right)} $ in relations
(\ref{eq22}) correspond to those classical values what means that
optimal cloning procedure (see Exp.~(\ref{eq20})) introduces some
error to both of transmitted and stored field. At the same time
the third correlation coefficient $C_{3}^{\left( {X} \right)} $ in
Exp.~(\ref{eq22}) demonstrates essentially non-classical
correlation between two clones at the output of cloning device in
Fig.4.

Last time the fidelity criterion is used to describe the
processing of quantum information -- see e.g.
Refs.\refcite{braunstein}--\refcite{grosshans},\refcite{vanloock}.
For optimal and symmetric cloning procedure of coherent optical
field the fidelity $F$ can be evaluated as:

\begin{equation}
\label{eq18new} F=\frac{2}{\sqrt{\left( 2+V_Q\right) \left(
2+V_P\right)}},
\end {equation}

\noindent where $V_Q=V_Q^{out}-V_Q^{in}$, and
$V_P=V_P^{out}-V_P^{in}$ are variances that correspond to the
noises introduced by cloning procedure. For two identical clones
the magnitude of fidelity is $F=2/3$.

Let us now examine the optimal cloning procedure by
Eqs.~(\ref{eq19}), (\ref{eq20}) for the BEC state (\ref{eq16}) of
the EIT-medium.

In the BEC state (\ref{eq16}) under the EIT condition for the mean
values $\left\langle {X_{f,\xi} ^{out}}  \right\rangle $ and for
the variances $V_{X,f}^{out} $ of output polariton parameters we
obtain:

\begin{equation}
\label{eq24} \left\langle {Q_{f,\xi} ^{out}}  \right\rangle =
\left\langle {Q_{f}^{in}} \right\rangle \mp \sqrt {2N} \left|
{\alpha _{1}}  \right|\left| {\alpha _{2}}  \right|\cos\varphi ,
\end{equation}

\begin{equation}
\label{eq25} \left\langle {P_{f,\xi} ^{out}}  \right\rangle =
\left\langle {P_{f}^{in}} \right\rangle \mp \sqrt {2N} \left|
{\alpha _{1}}  \right|\left| {\alpha _{2}}  \right|\sin\varphi ,
\end{equation}

\begin{equation}
\label{eq26} V_{Q,f}^{out} = V_{Q,\xi} ^{out} = V_{Q,f}^{in} + 1 -
2\left| {\alpha _{1}} \right|^{2}\left| {\alpha _{2}}
\right|^{2}\cos^{2}\varphi ,
\end{equation}

\begin{equation}
\label{eq27} V_{P,f}^{out} = V_{P,\xi} ^{out} = V_{P,f}^{in} + 1 -
2\left| {\alpha _{1}} \right|^{2}\left| {\alpha _{2}}
\right|^{2}\sin^{2}\varphi.
\end{equation}

The expressions (\ref{eq24})--(\ref{eq27}) demonstrate the
deviations from ``usual'' optimal cloning procedure -- cf. with
Eqs.~(\ref{eq21a}),(\ref{eq21b}). They are extremal (maximal) for
normalized atomic populations $|\alpha_1|^2=|\alpha_2|^2=1/2$ and
relative phase $\varphi=\pi/4$. Formally, using an expressions
(\ref{eq19}), (\ref{eq20}) for relevant correlation coefficients
(\ref{eq6})--(\ref{eq8}) and fidelity parameter $F$ defined in
Eq.~(\ref{eq18new}) one can obtain extremal values $C_{1,2}^{(Q)}
= C_{1,2}^{(P)} = 4/7$, $C_3^{(Q)}=C_3^{(P)}=25/49$, $F=0.8$.
However for good quantum cloning procedure we should require the
similarity for output and input quadrature components for the
scheme in Fig.4 which is possible exactly in adiabatic
approximation (\ref{eq3}). In particular, in the strong limit when
$\left| {\alpha _{1}} \right| \to 0$ and $\left| {\alpha _{2}}
\right| \simeq 1$ respectively, we obtain the same results as in
Exp.~(\ref{eq22}) with fidelity $F=2/3$.

In the paper we are not consider the problem of decoherence of
atomic and/or optical system. Although ultracold three level
condensate medium under the EIT condition is robust in respect of
absorption and Doppler broadening in single pass regime (see e.g.
Ref.~\refcite{liu}) the decoherence and losses become important
especially for long lived quantum memory purposes. Recently
various multi-pass protocols have been proposed for retrieval of
quantum optical state, verification of quantum storage state and
for increasing the fidelity,
respectively\cite{julsgaard}$^,$\cite{sherson}. Indeed the
retrieval procedure for continuous variables consists of mapping
of measured light quadrature back into the atomic quadtrature
variable of the medium with certain feedback gain. In some sense
such a technique is similar to the methods proposed early to
achieving good QND-measurement of light quadrature component or
photon number in signal field -- see e.g.
Ref.~\refcite{braginsky}. The problem of retrieval procedure for
the schemes proposed by us in Fig.2 and Fig.4 require detail
analysis what is not subject of consideration in the paper.
However it is important to note that multi-pass protocols
obviously are limited by the lifetime of atomic system what in our
case of condensate medium is large enough (up to few seconds --
see e.g.
Refs.~\refcite{liu},\refcite{prokhorov1},\refcite{anglin}.

In Appendix we pay our attention to the problem of stimulated Raman
photoassociation process as an real way for possible experimental
observation of the phenomena under consideration (in Figs.3,4).

\section{Conclusion}

In the present paper an opportunity to store quantum state of
optical field into the bright and dark polaritons arising in the
EIT-like medium is demonstrated. We propose two different schemes
for realization of quantum memory for photons. One possibility is
connected with the QND recording of two Hermitian quadratures of
signal optical field. The relevant correlation coefficients
demonstrate possibility to achieve quantum level of information
storage in the BEC medium. Another possibility is represented by
quantum cloning procedure for signal field in original cloning
device offered by us. In last case presence of the EIT effect in
condensate medium is absolutely necessary for the optimal cloning
procedure under the adiabatic approximation condition.

The physical schemes proposed by us can be useful in problem of
quantum cryptography for the first time. Recently, various quantum
key distribution (QKD) protocols have been proposed for continuous
variables\cite{cerf}$^,$\cite{grosshans}$^,$\cite{silberhorn}$^,$\cite{weedbrook}.
In particular, it have been shown (see e.g. Ref.~\refcite{cerf})
that for individual eavesdropping by beam splitter or cloning
machines (when we have less than 3~dB total losses in the channel)
these protocols are secure whenever $I_{AB} > I_{AE}$ ($I_{AB(E)}$
is information rate between Alice and Bob (Eve)) with the usage of
special reconciliation and privacy amplification procedure.
However, in some cases it is possible to create secure key beyond
the 3~dB limit when $I_{AB}<I_{AE}$ (or $I_{AB}<I_{BE}$) with the
help of appropriate postselection
procedure\cite{silberhorn}$^,$\cite{weedbrook}. In general the
information rates for Gaussian channels depends on signal-to-noise
ratio and, more precisely, on the variance of quadrature
components measured by Eve. From our point of view the schemes
proposed by us in Fig.2 and Fig.4 can be used by Eve for
individual attacks on the communication channel. In this case Eve,
for example, can explore simultaneous spin squeezing of the
"quadratures" $Q_\xi^{out}$ and $P_\xi^{out}$ (see
Eqs.~(\ref{eq26}), (\ref{eq27})) when she perform measurements
with her clone. The detailed consideration of this problem as well
as security of QKD problem in this case is subject of separate
analysis.

\section*{Acknowledgments}

This work was supported by the Russian Foundation for Basic
Research (grants No 04-02-17359 and 05-02-16576) and some Federal
programs of the Russian Ministry of Science and Technology. We are
grateful to Prof. Michele Leduc and her collaborators Steven Moal,
Maximilien Portier from Kastler Broseel Laboratory for fruitful
discussions on the Raman photoassociation subject and to Prof.
Helmuth Ritsch for introduction in his paper\cite{winkler}. We
also acknowledge to Referee for valuable comments and to Dr.
Andrey Leksin for assistance in preparation of the paper. One of
the authors (A.P. Alodjants) is grateful to the Russian private
Found ``Dynasty'' and International Center for Theoretical Physics
in Moscow for financial support.

\appendix{Stimulated Raman two-color photoassociation under the EIT condition}

Here we analyze the process of stimulated Raman two-photon photoassociation
in the presence of the EIT effect. The approach to the problem is based on
the stimulated Raman adiabatic passage (STIRAP) procedure resulting in
coherent coupling of the long life time levels in free atoms and as a
result, in establishment of molecular states. The process can be described
in formalism of the $\Lambda $-scheme in Fig.1.

In fact, initially we have a free pair of the BEC atoms in the
level $\left| {1} \right\rangle $, determined by energy $2E_{1} $.
Then, control field (frequency $\omega _{1} $) creates the
molecules in the excited molecular state $\left| {3} \right\rangle
$, i.e. described by energy $E_{3} $. Another laser field
(frequency $\omega _{2} $) removes adiabatically the molecules to
the lower state $\left| {2} \right\rangle $ of molecular
condensate. Thus, the STIRAP procedure is very closed physically
to the EIT observation in three level atomic BEC (cf.
Ref.~\refcite{liu} with
Refs.~\refcite{winkler},\refcite{koelemeij}).

To describe the process let us suppose that the signal field is a
quantum field characterizing by annihilation photon operator $f$.
In the three coupled condensate modes approximation the
Hamiltonian for stimulated Raman photoassociation can be
represented in the form\cite{drummond}:

\begin{eqnarray}
\nonumber H & = & \sum\limits_{i = 1,2,3} {E_{i}} a_{i}^{ +} a_{i}
+ \frac{{\hbar }}{{2}}\sum\limits_{ij = 1,2,3} {\lambda _{ij}}
a_{i}^{ +} a_{j}^{ +} a_{i} a_{j}\\ {} & - & \frac{{\hbar}
}{{2}}\left( {\kappa e^{ - i\omega _{1} t}a_{3}^{ +} a_{1}^{2} +
\Omega _{2} e^{ - i\omega _{2} t}a_{3}^{ +} a_{2}^{} f + H.c.}
\right) \label{eqA1}
\end{eqnarray}

\noindent where $a_{j} $ is the $j$-th annihilation operator for
atoms ($j = 1$), molecules ($j = 2$) and excited molecules ($j =
3$), respectively; the coefficients $\lambda _{ij} $ characterize
the atom-atom, atom-molecule and molecule-molecule scattering in
the Born approximation; the parameters $\kappa $ and $\Omega _{2}
$ specify the Rabi frequencies. Then, in rotating frame (realized
by substitution $a_{j} \to a_{j} \exp\left( { - {{\left( {E_{j} +
\Delta _{j}}  \right)t} \mathord{\left/ {\vphantom {{\left( {E_{j}
+ \Delta _{j}}  \right)t} {\hbar} }} \right.
\kern-\nulldelimiterspace} {\hbar} }} \right)$) the mean-field
equations have the form:

\begin{eqnarray}
\label{eqA2} \frac{{d\left\langle {a_{1}}  \right\rangle} }{{dt}}
& = & - \gamma _{1} \left\langle {a_{1}}  \right\rangle + i\delta
_{1} \left\langle {a_{1}} \right\rangle + i\Omega _{1}^{\ast}
\left\langle {a_{3}}  \right\rangle ,\\ \label{eqA3}
\frac{{d\left\langle {a_{2}}  \right\rangle} }{{dt}} & = & -
\gamma _{2} \left\langle {a_{2}}  \right\rangle + i\delta _{2}
\left\langle {a_{2}} \right\rangle + i\frac{{\Omega _{2}^{}}
}{{2}}\left\langle {f^{ +} } \right\rangle \left\langle {a_{3}}
\right\rangle,\\ \label{eqA4} \frac{{d\left\langle {a_{3}}
\right\rangle} }{{dt}} & = & - \gamma _{3} \left\langle {a_{3}}
\right\rangle + i\frac{{1}}{{2}}\Omega _{1} \left\langle {a_{1}}
\right\rangle + i\Omega _{2}^{} \left\langle {f} \right\rangle
\left\langle {a_{2}}  \right\rangle ,\\ \label{eqA5}
\frac{{d\left\langle {f} \right\rangle} }{{dt}} & = &
i\frac{{\Omega _{2} }}{{2}}\left\langle {a_{2}^{ +} }
\right\rangle \left\langle {a_{3}} \right\rangle,
\end{eqnarray}

\noindent
where $\delta _{j} = \Delta _{j} - \sum\limits_{k = 1}^{3} {\lambda _{jk}
\left\langle {a_{k}^{ +}  a_{k}}  \right\rangle}  $ is effective frequency
shift, $2\Delta _{1} = {{\left( {E_{3} - 2E_{1}}  \right)} \mathord{\left/
{\vphantom {{\left( {E_{3} - 2E_{1}}  \right)} {\hbar} }} \right.
\kern-\nulldelimiterspace} {\hbar} } - \omega _{1} $, $\Delta _{2} =
{{\left( {E_{3} - E_{2}}  \right)} \mathord{\left/ {\vphantom {{\left(
{E_{3} - E_{2}}  \right)} {\hbar} }} \right. \kern-\nulldelimiterspace}
{\hbar} } - \omega _{2} $ are detunings in rotating frame; $\gamma _{j} $
defines spontaneous decay rate from $j$th level. Neglecting by nonlinear
character of atom-molecular interaction for free-bound transition we
introduce the effective Rabi frequency $\Omega _{1} \simeq \kappa
\left\langle {a_{1}}  \right\rangle $.

So, set of Eqs.~(\ref{eqA2})--(\ref{eqA4}) characterize the usual
STIRAP procedure for two-color photoassociation
effect\cite{drummond}$^,$\cite{winkler}. Last Eq.~(\ref{eqA5})
takes into account the properties of signal field under the STIRAP
procedure realization that absolutely necessary in the EIT scheme.

Alternatively, set of Eqs.~(\ref{eqA2})--(\ref{eqA5}) can also be
derived directly from the interaction Hamiltonian :

\begin{equation}
\label{eqA6} H = - \hbar \delta _{1} a_{1}^{ +}  a_{1} - \hbar
\delta _{2} a_{2}^{ +} a_{2} - \frac{{\hbar} }{{2}}\left( {\Omega
_{1} a_{3}^{ +}  a_{1}^{} + \Omega _{1}^{\ast}  a_{3}^{ +}
a_{1}^{}}  \right) - \frac{{\hbar \Omega _{2}} }{{2}}\left(
{a_{3}^{ +}  a_{2}^{} f + f^{ +} a_{2}^{ +}  a_{3}^{}} \right),
\end{equation}

\noindent that obviously can be associated with the EIT problem in
$\Lambda $-systems\cite{fleischauer}$^,$\cite{prokhorov1}.
Therefore, we can now express the oblique angle $\theta $ in the
terms of molecular Rabi frequencies according to expression
(\ref{eq2}).

Thus, the bright and dark polaritons are specified in this case by
collective phenomenon of the atom-molecular and optical field
excitation. The experimental realization of such a stimulated
Raman two color photoassociation is difficult, in principle, due
to severeal physical reasons (see e.g. Ref.~\refcite{koelemeij}).
But demonstration of dark state resonance performed recently in
Ref.~\refcite{winkler} is an important step toward the EIT-effect
observation.


\section*{References}
\begin{thebibliography}{0}

\bibitem{julsgaard} B. Julsgaard, J. Sherson, J. I. Cirac, J. Fiurasek, E. S. Polzik,
{\it Nature} {\bf 432,} 482 (2004).

\bibitem{kozhekin} A. E. Kozhekin, K. M\o lmer, E. Polzik,
{\it Phys. Rev}. {\bf A62,} 033809 (2000).

\bibitem{fleischauer} M. Fleischauer, M. D. Lukin,{ \it Phys. Rev.}
{\bf A65,} 022314 (2002).

\bibitem{liu} C. Liu, Z. Dutton, C.H. Behroozi, L.V. Hau, {\it Nature,}
{\bf 409,} 490 (2001).


\bibitem{prokhorov1} A. V. Prokhorov, A. P. Alodjants, S. M. Arakelian, {\it JETP
Letts}, {\bf 80,} 739 (2004).

\bibitem{phillips} D. F. Phillips, A. Fleischhauer, A. Mair, R.
L. Walsworth, M. D. Lukin, {\it Phys. Rev. Letts.} {\bf 86,} 783
(2001).

\bibitem{fiurasek} J. Fiur\'asek, N. J. Cerf , E. S. Polzik,
{\it Phys. Rev. Letts.} {\bf 93,} 180501 (2004).

\bibitem{braunstein} L. Braunstein, N. J. Cerf, S. Iblisdir, P. Van Loock, S.
Massar, {\it Phys. Rev. Letts.} {\bf 86,} 4438 (2001).

\bibitem{cerf} N. J. Cerf, S. Iblisdir, G. Van Assche, {\it Europ.
Phys. J.} {\bf D 18,} 211 (2002).

\bibitem{grosshans} F. Grosshans, P. Grangier, {\it Phys. Rev.} {\bf A64,} 010301-1 (2001); {\it Phys. Rev.
Letts.} {\bf 88,} 057902-1 (2002).

\bibitem{braginsky} V. B. Braginsky, F. Ya. Khalili, {\it Rev. of Mod.
Phys.} {\bf 68,} 1 (1996).

\bibitem{alodjants} A. P. Alodjants, S. M. Arakelian, A. S. Chirkin, {\it Appl.
Phys.} {\bf B53,} 65 (1998).

\bibitem{drummond} P. D. Drummond, K. V. Kheruntsyan, D. J. Heinzen, R. H. Wynar,
{\it Phys.Rev.} {\bf A65,} 063619-1 (2002).

\bibitem{winkler} K. Winkler, G. Thalhammer, M. Theis, H. Ritsch,
R. Grimm, J. Hecker Denschlag, {\it Phys. Rev. Letts.} {\bf 95,}
063202-1 (2005).


\bibitem{anglin} J. R. Anglin, A. Vardi, {\it Phys. Rev.} {\bf
A64,} 013605-1 (2001).

\bibitem{adams} C. S. Adams, M. Sigel, J. Mlynek, {\it Phys. Rep.}
{\bf 240,} 143 (1994).

\bibitem{levy} L.-P. Levy, {\it Magnetism and Superconductivity}
(Springer-Verlag, Berlin, 2000), p.467.

\bibitem{prokhorov2} A. V. Prokhorov, A. P. Alodjants, S. M. Arakelian,
{\it Optics and Spectr.} {\bf 94,} 50 (2003).

\bibitem{cirac2} J. I. Cirac, M. Lewenstein, K. Molmer, P.
Zoller, {\it Phys. Rev.} {\bf A57,} 1208 (1998).

\bibitem{vanloock} P. Van Loock, S. L. Braunstein, H. J. Kimble,
{\it Phys. Rev.} {\bf A62,} 022309-1 (2000).

\bibitem{sherson} J. Sherson, A. S. S\o rensen, J. Fur\'asek, K.
M\o lmer, E. S. Polzik, "Light qubit storage and retrieval using
macroscopic atomic ensembles", quant-ph/0505170.

\bibitem{silberhorn} Ch. Silberhorn, T. C. Ralph, N. Lutkenhaus,
G. Leuchs {\it Phys. Rev. Letts.} {\bf 89,} 167901-1 (2002).


\bibitem{weedbrook} A. M. Lance, T. Symul, V. Sharma et al. {\it Phys.
Rev. Letts.} {\bf 95,} 180503-1 (2005).

\bibitem{koelemeij} J. C. J. Koelemeij, M. Leduc, {\it Eur. Phys. J.}
{\bf D31,} 263 (2004).


\end{thebibliography}
\end{document}